\documentclass{moriond}

\def\Journal#1#2#3#4{{#1} {\bf #2}, #3 (#4)}

\def\be{\begin{equation}}
\def\ee{\end{equation}}
\def\bea{\begin{eqnarray}}
\def\eea{\end{eqnarray}}

\usepackage{subcaption}
\usepackage{enumitem}
\usepackage{textgreek}

\usepackage{blindtext}
\usepackage{scrextend}
\addtokomafont{labelinglabel}{\sffamily}

\newcommand{\Photo}{}

\begin{document}
\vspace*{4cm}

\title{NEWS-G\\
Light dark matter search with a Spherical Proportional Counter\\
First results and Future prospects~\footnote{Presented at 53$^{\rm rd}$ Rencontres de Moriond on Electroweak Interactions and Unified Theories.}
}
\author{Ioannis Katsioulas\\on behalf of the NEWS-G collaboration}
\hfill\break
\address{IRFU, CEA, Universit\'e Paris-Saclay,\\ F-91191 Gif-sur-Yvette, France   }

\maketitle\abstracts{
NEWS-G (New Experiments With Spheres-Gas) is an experiment aiming to shine a light on the dark matter conundrum using a novel gaseous detector, the Spherical Proportional Counter. NEWS-G uses light noble gases, such as hydrogen, helium, and neon, as targets, to search for light dark matter down to the sub-GeV/c${}^{2}$ mass region. The first detector of NEWS-G, is a 60 cm diameter sphere already operated in the Underground Laboratory of Modane, while the full-scale detector, 140 cm in diameter, will be installed in SNOLab at the end of this year. In this work, we present the first NEWS-G results with neon as target, which excludes at 90$\%$ confidence level cross-sections above $4.4\cdot 10{}^{37}$ cm${}^{2}$ for a candidate with a mass 0.5 GeV$/$c${}^{2}$ based on 9.7 kg$\cdot$days of exposure. The status of the project and prospects for the future are also discussed.
}

\section{Introduction}

For over 80 years, since the first observations by F.~Zwicky, Dark Matter (DM) constitutes a basic pillar of the cosmological model of our Universe and although there are compelling evidence in all astrophysical scales, the particle content of DM remains elusive. From astrophysical observations we know that DM should be ``cold", non-baryonic, weakly interacting with a density of $\Omega h^{2}=0.1186\pm 0.002$~\cite{pdg}. A plethora of particles has been proposed over the years that match these criteria, none predicted by the Standard Model. This generic class of DM candidates is known as Weakly Interacting Massive Particles (WIMPs). Initially, WIMPs in the 10 GeV/c$^{2}$ -- 1 TeV/c${}^{2}$ mass range were favored by supersymmetric models. But the non-findings of passive experiments and at the LHC motivate searches in masses below 10 GeV/c${}^{2}$, where many new theoretical approaches such as the asymmetric dark model and dark sector predict DM candidates.
Direct DM detection in this mass range is beyond the capabilities of the conventional experiments based on dual phase TPCs and solid state detectors which use heavy elements such as Xe, Ge, Ar or Si as targets due to two main reasons. First, is the kinematic issue i.e. it is less likely to produce energetic recoils with light WIMP scattering on heavy targets because of the minimum WIMP relative velocities required that are scarce in our Galaxy. The second issue is that heavy recoils have a large ionization deficit that is described by the ionization quenching factor~\cite{Santos}. For example a Xe recoil with kinetic energy equal to 500 eV will induce in a medium of the same type will deposit $\sim50$ eV through ionization. As a result, it would require a detector with a energy threshold at the single ionization electron to be detected. Both the kinematic and ionization issues are much less prominent for lighter targets, making them more appropriate for the detection of light DM candidates. In this work, we present the concept of the NEWS-G experiment which by using an innovative gaseous detector, the Spherical Proportional Counter (SPC) filled with Ne, He and H gas mixtures as targets, aiming to search for light WIMPs with unprecedented sensitivity.

\section{The Spherical Proportional Counter}

The SPC is a spherical gaseous detector~\cite{spc1}, presented in Fig.~\ref{fig:spc}, consisting of a grounded spherical shell which acts as the cathode and a small ball -the anode- placed at the center of the sphere, supported by a grounded metallic rod, to which the high voltage (HV) is applied and from which the signal is read-out (the sensor). The electric field, which is given by
$$E(r) = \frac{r_{A}r_{C}}{r_{C}-r_{A}} \frac{V_{0}}{r^{2}}$$
where $r_{A}$ is the anode radius, $r_{C}$ the cathode radius, $V_{0}$ the voltage applied on the anode and $r$ the distance from the center of the detector. The dependence on the inverse of the detector radius squared results in the natural division of the detector into the drift volume where under the influence of the low electric field the electrons drift towards the anode until a few mm from the anode surface, i.e. the amplification region, where the charge multiplication takes place. 

The simplicity of the design permits the construction of a detector of this type solely by radiopure materials. A large detector can be constructed using tens of kgs of copper with the exception of parts of the sensor weighing less than one gram.

\begin{figure}
\centering
  \begin{subfigure}[b]{0.45\textwidth}
    \includegraphics[width=\textwidth]{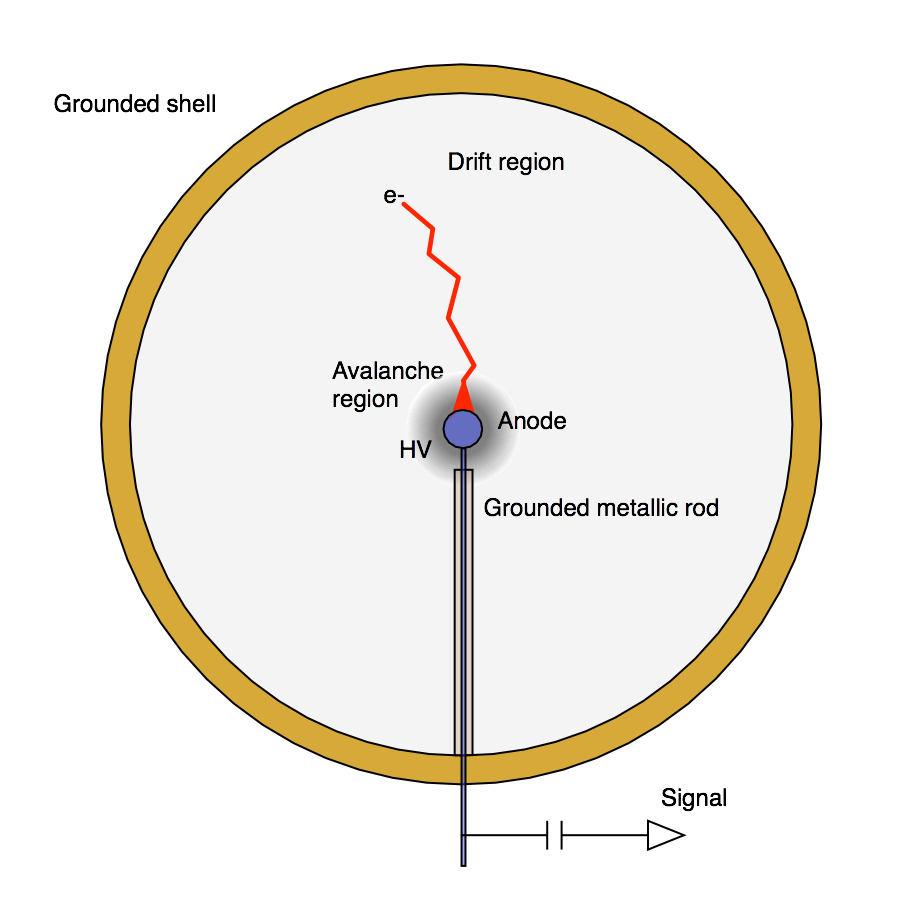}
    \caption{}
    \label{fig:spc1}
  \end{subfigure}
  \begin{subfigure}[b]{0.35\textwidth}
    \includegraphics[width=\textwidth]{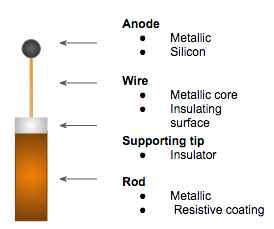}
    \caption{}
    \label{fig:spc2}
  \end{subfigure}
  
\caption[]{(a) SPC design and principle of operation and (b) illustration of the basic read out sensor \cite{recoils}. }
\label{fig:spc}
\end{figure}

The spherical geometry provides several advantages against other detector geometries, such as the parallel plate detectors or the cylindrical counters when it comes to building large detectors ($\sim$ 1 m${}^3$). The sphere has the lowest surface to volume ratio and it is the vessel that can easier sustain high pressures. In addition, it is the detector geometry that has the lowest capacitance compared to parallel plate or cylindrical detectors. The reason for this is that its capacitance, which is approximated by $C \simeq 4\pi \epsilon_{0} r_{A}$ when $r_{C}\gg r_{A}$, depends only on the radius of the anode and not on the size of the vessel, making it easily scalable. Also the high gain achieved with anode balls with order mm diameter make the detector construction more robust. All these advantages allow the development of large detectors with single electron detection threshold and increased signal to background ratio.

Another important advantage is that it provides the possibility to discriminate between events with long tracks and point like events and also provides information about the interaction radius. This is due to the difference in the collection time of the primary electrons in these events which is depicted in the rise time of the pulses~\cite{spc1,recoils}.

The symmetry of the spherical shape along with the possibility to reject events originating from a given radius provide a valuable handle to reject background which usually comes from the surface in rare event search experiments.

\section{The NEWS-G/LSM detector - First WIMP search run}
\begin{figure}
\centering
  \begin{subfigure}[b]{0.25\textwidth}
    \includegraphics[width=\textwidth]{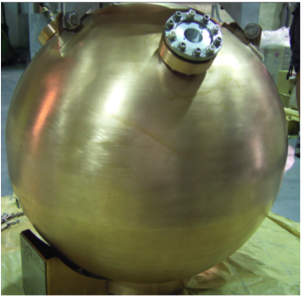}
    \caption{}
    \label{fig:sedine1}
  \end{subfigure}
  \begin{subfigure}[b]{0.15\textwidth}
    \includegraphics[width=\textwidth]{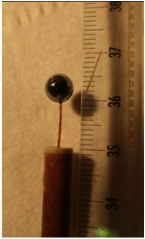}
    \caption{}
    \label{fig:sedine2}
  \end{subfigure}
  \begin{subfigure}[b]{0.36\textwidth}
    \includegraphics[width=\textwidth]{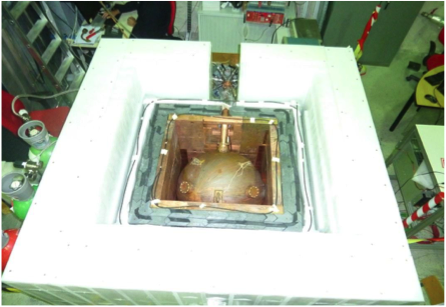}
    \caption{}
    \label{fig:sedine3}
  \end{subfigure}
\caption[]{(a) Picture of SEDINE, the 60 cm diameter prototype installed at LSM, (b) the sensor installed in SEDINE with a 6.3 mm in diameter Si ball and a 380 \textmu m diameter
insulated HV wire routed through a grounded copper rod and (c) the cubic shielding of SEDINE comprised of layer of copper, lead and polyethylene.}
\label{fig:spc}
\end{figure}

An SPC is already installed at the Laboratoire Souterrain de Modane (LSM) inside the Frejus tunnel, one of the deepest laboratories in the world, under a rock thickness of 4800 m water equivalent for protection against cosmic radiation. The detector, named SEDINE, consists of a 60 cm diameter sphere made of ultra pure (NOSV) copper (Fig.~\ref{fig:sedine1}) and a 6.3 mm diameter spherical sensor made of silicon located at the center of the vessel (Fig.~\ref{fig:sedine2}). It is additionally protected from external radiation by a multi-layered cubic shielding composed of, from the inside to the outside, 8 cm of copper, 15 cm of lead and 30 cm of polyethylene (Fig.~\ref{fig:sedine3}). NEWS-G used SEDINE to perform its first search. The detector was filled with a mixture of neon (99.3\% in pressure) and methane (0.7\%) at a total pressure of 3.1 bar, corresponding to 280 g of target mass. The sensor was biased to 2520 V and a CANBERRA Model 2006 charge sensitive preamplifier was used to read-out the signal. The detector was operated under these conditions, in sealed mode for 42.7 days without interruption.
\subsection{Calibration}
The detector was calibrated using an ${}^{37}$Ar gaseous X-ray source and a ${}^{241}$Am--${}^{9}$Be neutron source. The ${}^{37}$Ar gas was added to the mixture of Ne:CH${}_{4}$ at the end of the run, providing a large sample of mono-energetic events at 2.82 keV and 270 eV from X-rays induced by electron capture in the K- and L-shells, respectively.  The ${}^{241}$Am--${}^{9}$Be neutron source was used to induce nuclear recoil events, homogeneously distributed in the detector volume down to the critical low energy range of 150 eVee - 250 eVee where our sensitivity to sub-GeV/c${}^{2}$ WIMPs is expected. This calibration is of great importance as it allows the detector response measurement in rise time for point-like energy depositions in the detector volume.

\begin{figure}
\centering
  \begin{subfigure}[b]{0.4\textwidth}
    \includegraphics[width=\textwidth]{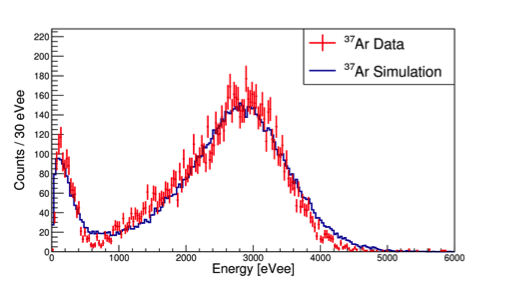}
    \caption{}
    \label{fig:calib1}
  \end{subfigure}
  \begin{subfigure}[b]{0.4\textwidth}
    \includegraphics[width=\textwidth]{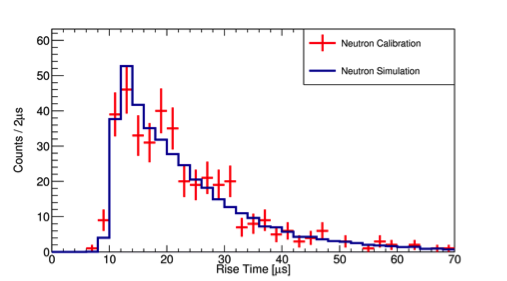}
    \caption{}
    \label{fig:calib2}
  \end{subfigure}
\caption[]{(a) Agreement between of ${}^{37}$Ar calibration data and simulation. The measured energy spectrum (red markers) exhibits two peaks at 2.82 keV and 270 eV from the electron capture in the K- and L-shell, respectively. The energy spectrum derived from the simulation (dark blue histogram) is scaled to the number of events recorded during the calibration run for comparison. (b) Comparison between the rise time distribution of events in the energy range 150 eVee - 250 eVee for events from neutron calibration data (red markers) and for simulated events with energy deposits homogeneously distributed in the volume (dark blue histogram) \cite{news}.}
\label{fig:calib}
\end{figure}

\section{Analysis of the LSM run and first NEWS-G results}

\subsection{Event simulation method - Background modeling}
For the determination of our sensitivity to light DM we simulated the detector response to ionizing radiation. In this simulation method we took into account all the basic processes that take place during the detector operation such as (a) the primary ionization for electronic recoils of energy $E_{R}$ with mean energy to produce an electron-ion pair $W_{\gamma}(E_{R})$ and for nuclear recoils with $W_{n}(E_{R}) = W_{\gamma}(E_{R})/Q(E_{R})$ where $Q(E_{R})$ is the SRIM derived quenching factor, (b) the electric field given by finite-element software after modeling the full detector geometry, (c) the drift and diffusion of primary charges which was parametrized using Magboltz coefficients, and (d) the avalanche process which was simulated using Garfield++, accounting for Penning transfers and fitting the results with a Polya distribution \footnote{$ P\Bigg( \frac{n}{\langle n \rangle} \Bigg) = \frac{(1+\theta)^{(1+\theta)}}{\Gamma(1+\theta)}\Bigg( \frac{n}{\langle n \rangle} \Bigg)^{\theta}exp\Bigg[ -(1+\theta) \frac{n}{\langle n \rangle} \Bigg]$, where $\langle n \rangle$ is the mean gain and $\theta$ the shape parameter}. 

The final pulse for an event was constructed by adding the contributions of all PEs that reached the sensor. To these pulses we also added a baseline randomly chosen from empty pre-traces of events recorded during the physics-run to account for realistic noise. The simulated pulses were processed with the same algorithms as real pulses.

The simulation method was validated by comparing the simulation to the results of the ${}^{37}$Ar calibration with the measurements which we present at (Fig.~\ref{fig:calib1}). We can see the good agreement but also the reproduction of the non-Gaussianity of the 2.82 keV line caused by the field anisotropy in the avalanche region. In Fig.~\ref{fig:calib2} we present also the comparison of the rise time distribution of the neutron induced nuclear recoils during the ${}^{241}$Am--${}^{9}$Be calibration and the simulated nuclear recoils. The good agreement between the two distributions validates the Magboltz derived drift parameters used in the simulation to calculate the rise time of the pulses. 

Using this method we modeled our background which is categorized into surface and volume events. The surface events are uniformly distributed in the vessel and originate from radon daughter decays. Volume events originate from high energy \textgamma-rays from ${}^{208}$Tl and ${}^{40}$K present in the rock, and from the decay chains of ${}^{238}$U and ${}^{232}$Th contained in the copper shell and the shielding itself.  The results of these simulations have shown that the pulse rise time provides useful statistical discrimination against surface events down to our analysis threshold of 150~eVee.

\subsection{Data analysis}
A conservative analysis threshold was set at 150 eVee, above the trigger threshold of $\sim$ 36 eVee, to ensure a $\sim$ 100 \% trigger efficiency. Further on, a quality cut was applied to reject events within a 4 second window after triggering to remove non-physical events thus introducing a dead time resulting to 20.1\% exposure loss. After the quality cut we defined a preliminary Region Of Interest (ROI) in rise time 10 \textmu s - 32 \textmu s and energy 150 eV - 4000 eV. Side band regions were used to determine the expected backgrounds in the preliminary ROI. The event rate measured in the 4000 eVee - 6000 eVee energy range was used to extrapolate the expected Compton background down to lower energy assuming a flat recoil energy spectrum. The side band region 
was used together with our simulation to extrapolate the expected number and distribution of surface events leaking in the preliminary ROI. For a total exposure of 34.1 live-days$\times$0.283 kg = 9.7 kg$\cdot$days, 1620 events were recorded in the preliminary ROI.

The ROI was tuned by using a Boosted Decision Tree (BDT); a machine learning algorithm which was trained with $10^{5}$ simulated background events and signal events for 8 different WIMP masses (from 0.5 to 16 GeV/c$^{2}$). For each WIMP mass, the events where classified as background-like or signal-like using the BDT score (ranging  between -1 and 1 respectively) resulting in a WIMP-mass-dependent fine-tuned ROI in the rise time vs. energy plane. Details about the principle of the analysis and the simulations can be found in the recent NEWS-G~\cite{news}.

\subsection{Results}

\begin{figure}
\centering
\includegraphics[width=0.6\textwidth]{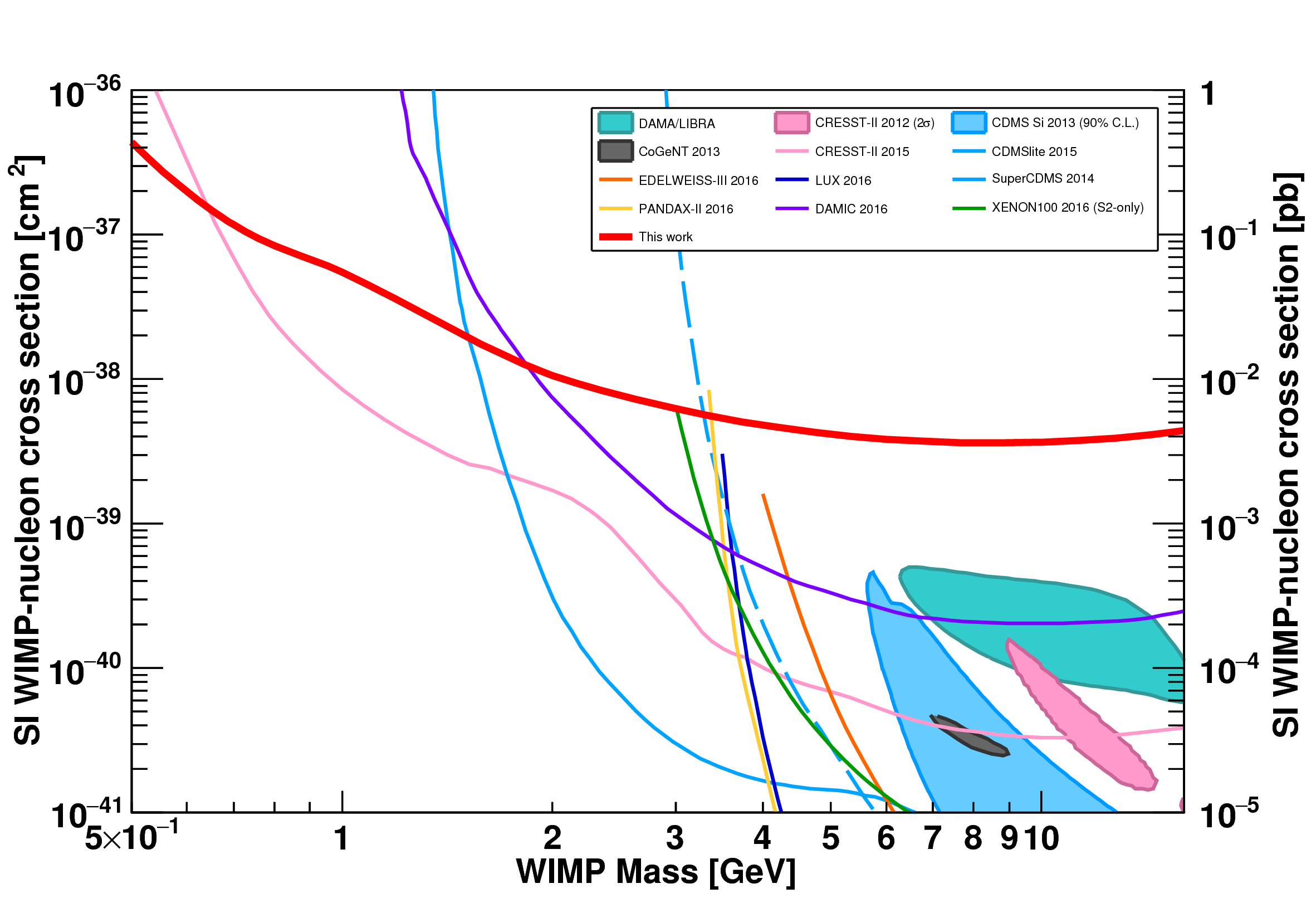}
\caption[]{Constraints in the Spin-Independent WIMP-nucleon cross section versus WIMP mass plane. The result from this analysis is shown in solid red~\cite{news}.}
\label{fig:news}
\end{figure}

 For each WIMP mass and considering as candidates all the events observed in the corresponding fine-tuned ROI, a 90 \% Confidence Level (C.L.) upper limit on the spin-independent WIMP-nucleon scattering cross section was derived using Poisson statistics. The recoil energy spectrum used to derive the sensitivity to light DM is based on standard assumptions of the DM-halo model\footnote{dark matter density $\rho_{DM} = 0.3\;{\rm GeV}/c^{2}/{\rm cm}^{3}$, galactic escape velocity of $v_{\rm esc} = 544\;{\rm km/s}$, asymptotic circular velocity of $v_{0} = 220\;{\rm km/s}$}. The resulting exclusion limit is presented at Fig.~\ref{fig:news} as a solid red line, setting new constraints on the spin-independent WIMP-nucleon scattering cross-section below 0.6 GeV/c$^{2}$ and excludes at 90\% C.L. a cross-section of $4.4\times10^{-37}$ cm${}^{2}$ for a 0.5 GeV/c$^{2}$ light DM candidate mass.

\section{Future prospects - Conclusions}
 With the first results of NEWS-G we demonstrated the high potential of Spherical Proportional Counters being used for the search of low-mass DM. A new physics run is under way by operating SEDINE filled with He based gas mixture and a new physics run with H rich gas mixtures is envisioned. These runs will allow for more optimized momentum transfers for low-mass particles in the GeV/c$^{2}$ mass range, and increase our sensitivity to sub-GeV/c$^{2}$ light DM. The next phase of the experiment will build upon the experience acquired from the operation of the SEDINE prototype at the LSM. It will consist of a 140 cm diameter sphere, made of extremely low activity copper (in the range of a few \textmu Bq/kg of U and Th impurities) that is going to begin operation in SNOLab by summer 2019. The design for the shielding of the new sphere is more compact and advanced. It consists of shell of 22 cm of low activity lead lined with 3 cm of archaeological lead, inside a 40 cm thick polyethylene shield. The new setup will include dedicated handling to avoid radon entering the detector at any time. With such improvements we expect a significant reduction of the backgrounds levels, relative to the above results, and allow sensitivity down to cross sections of $\mathcal{O}$ $({10}^{-41}$ cm$^{2}$). The use of H and He targets will allow us to reach WIMP mass sensitivity down to 0.1 GeV/c$^{2}$. Finally, NEWS-G is developing novel detector instruments to facilitate the efficient operation of large detectors, such as the NEWS-G/SNO detector that may be influenced by the especially low electric field in the far radii of the detector volume. For this reason, new multi-ball sensors (ACHINOS~\cite{achinos}) are being developed (Fig.~\ref{fig:achinos1}) that are composed of multiple anode balls equidistantly placed on a virtual spherical surface and all biased at the same potential. Their collective influence results in an increased  field magnitude at large radii while maintaining the ability to reach high gain operation provided by anode diameter in the order of a mm. Developments such as this open the way for even larger detectors and operation in higher pressure.
\begin{figure}
\centering
  \begin{subfigure}[b]{0.18\textwidth}
    \includegraphics[width=\textwidth]{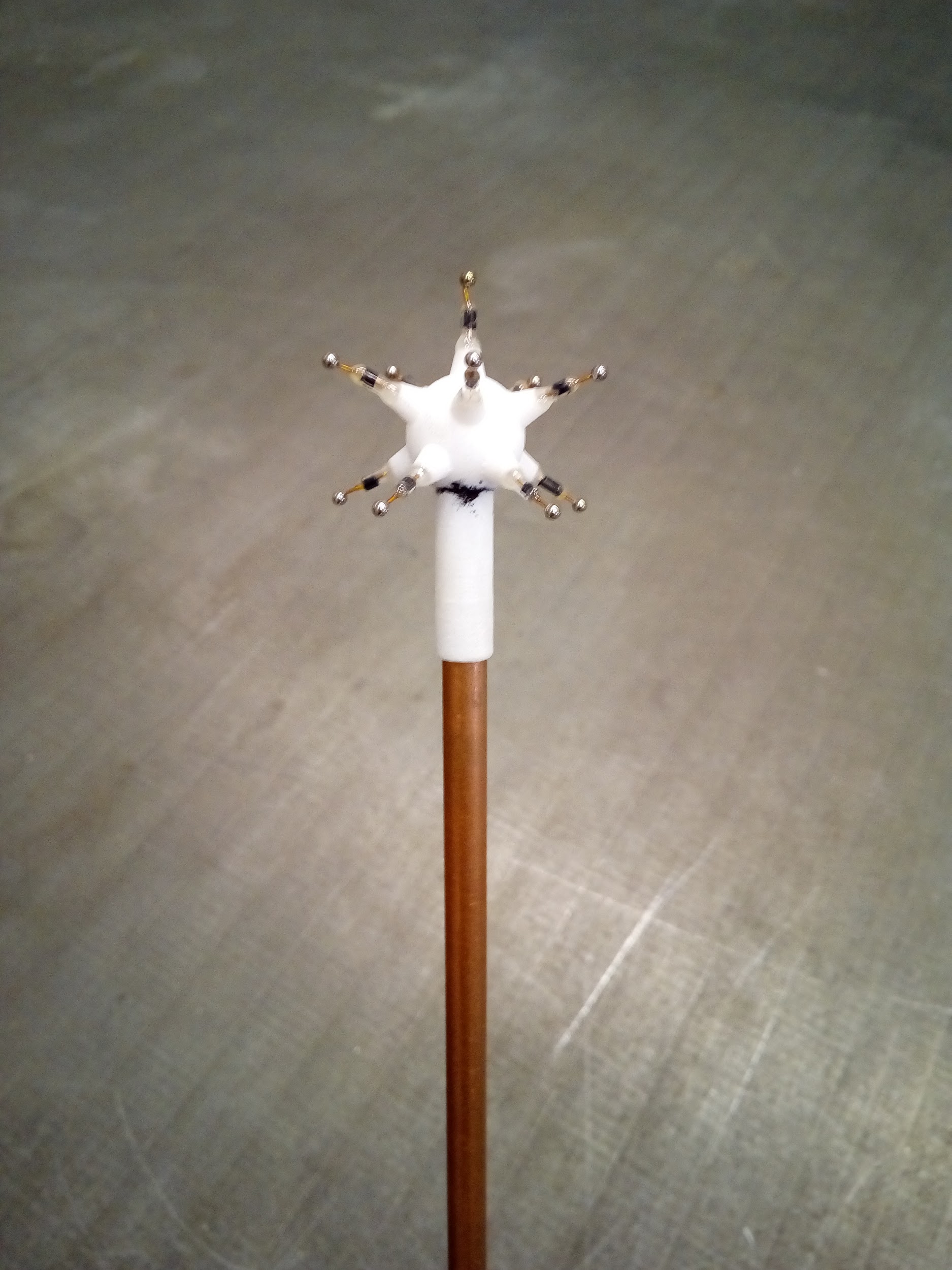}
    \caption{}
    \label{fig:achinos1}
  \end{subfigure}
  \begin{subfigure}[b]{0.35\textwidth}
    \includegraphics[width=\textwidth]{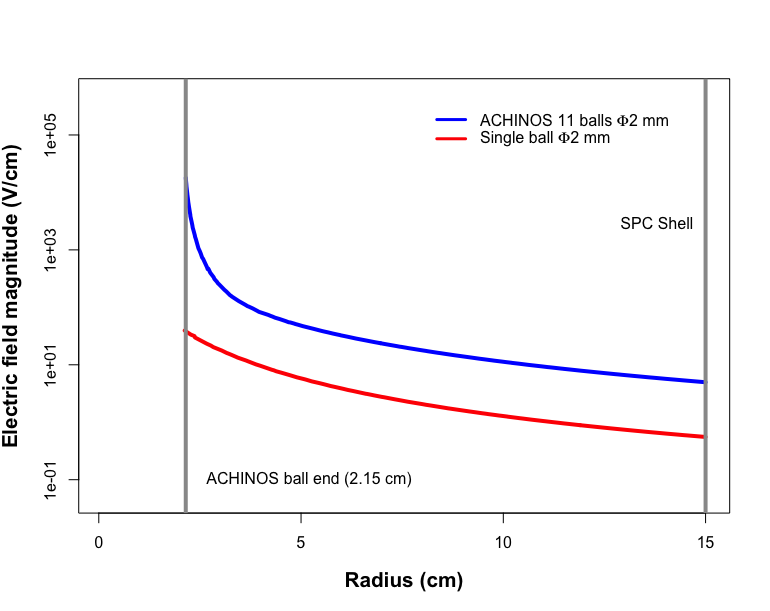}
    \caption{}
    \label{fig:achinos1}
  \end{subfigure}
\caption[]{(a) An ACHINOS prototype with 11 balls of 2 mm in diameter constructed using 3D printed materials (b) the electric field magnitude versus the radius in the case of a single 2-mm in diameter anode ball and in the case of the 11-ball ACHINOS sensor distributed on a 36-mm diameter sphere both placed in the center of a 300-mm diameter detector. The electric field in the far radii is $\sim9$ times higher for a detector equipped with an ACHINOS sensor than a single ball sensor placed at the same bias~\cite{achinos}.}
\label{fig:achinos}
\end{figure}
\section{Acknowledgments}
This work is funded by the French National Research Agency (ANR-15-CE31-0008).
\newpage
\section*{References}


\begin{thebibliography}{99}

\bibitem{pdg} C.~Patrignani {\it et al}, \Journal{Chin. Phys. C}{40}{ 100001}{2017}.

\bibitem{Santos} D.~Santos {\it et al}, \Journal{arXiv}{astro-ph}{0810.1137}{2008}.

\bibitem{spc1} I.~Giomataris {\it et al}, \Journal{Journal of Instrumentation}{3}{09}{2008}.

\bibitem{news} Q.~Arnaud  {\it et al}, \Journal{Astroparticle Physics}{97}{54 – 62}{2018}.

\bibitem{recoils}I.~Savvidis {\it et al}, \Journal{ Nuclear Instruments and Methods in Physics Research. Section A}{877}{220 - 26}{2018}.

\bibitem{srim}J.F.~Ziegler {\it et al}, \Journal{ Nuclear Instruments and Methods in Physics Research. Section B}{268}{1818 - 1823}{2010}.

\bibitem{achinos}A.~Giganon {\it et al}, \Journal{Journal of Instrumentation}{12}{P12031}{2017}.


\end{thebibliography}
\end{document}